# Measurements of the Rate Capability of Various Resistive Plate Chambers

M. Affatigato[a], U. Akgun[a], B. Bilki[b,d], F. Corriveau[e], B. Freund[b,e], N. Johnson[a], C. Neubüser[b,c], Y. Onel[d], J. Repond[b,*], and L. Xia[b]

[a] *Coe College,*
  *1220 1st Ave NE, Cedar Rapids, IA 52402, U.S.A.*
[b] *Argonne National Laboratory,*
  *9700 S. Cass Avenue, Argonne, IL 60439, U.S.A.*
[c] *DESY,*
  *Notkestrasse 85, D-22607 Hamburg, Germany*
[d] *University of Iowa,*
  *Iowa City, IA 52242-1479, U.S.A.*
[e] *McGill University,*
  *3600 University Street, Montreal, QC H3A2T8, Canada*

  E-mail: repond@anl.gov

ABSTRACT: Resistive Plate Chambers (RPCs) exhibit a significant loss of efficiency for the detection of particles, when subjected to high particle fluxes. This rate limitation is related to the usually high resistivity of the resistive plates used in their construction. This paper reports on measurements of the performance of three different glass RPC designs featuring a different total resistance of the resistive plates. The measurements were performed with 120 GeV protons at varying beam intensities.

KEYWORDS: Resistive Plate Chambers, rate capability.

---

[*] Corresponding author.

# Contents

## 1. Introduction

Resistive Plate Chambers (RPCs) were first introduced in the 1980's [1]. Their design typically features two (or more) resistive plates made of either Bakelite or glass. The readout board is placed on the outside of the chamber and contains strips or pads which pick up the signals inductively. RPCs are widely used in High Energy Physics experiments, foremost for triggering and precision timing purposes.

In this paper, we report on the measurement of the rate capability of RPCs. Due to the typically high resistance of the resistive plates used in the construction of RPCs, the devices exhibit a significant loss of efficiency, when subjected to high particle intensities [2]. The rate capability of glass RPCs based on three different designs, featuring different conductance per area of the glass plates in the range of $1 \times 10^{-12}$ to $6 \times 10^{-11}$ $\Omega cm^2$, was measured in a particle beam. Other techniques available to improve the rate capability, such as operation with reduced high voltage and signal threshold, are not explored in this paper. This work was performed in the context of studies of imaging calorimetry for a future lepton collider, as carried out by the CALICE collaboration [3].

## 2. Three different chamber designs

The rate capability of three different RPC designs was measured:

1) *2-glass RPCs with standard glass*
   The chambers were built with two standard soda-lime float glass plates with a thickness of 1.1 mm each. The gas gap was 1.2 mm. The chambers were $20 \times 20$ cm$^2$ in size. For more details on the chambers see [4].
2) *1-glass RPCs with standard glass*

The chambers were built with one standard soda-lime float glass plate with a thickness of 1.15 mm. The gas gap was also 1.15 mm. The size of the chamber was dictated by the size of the readout board, i.e. $32 \times 48$ cm$^2$. With only one glass plate the gas volume is defined by the glass plate and the anode board. Thus, the readout pads are located directly in the gas volume. For more details on the chamber, see [5].

3) *2-glass RPCs with semi-conductive glass*
These chambers utilize semi-conductive glass with a bulk resistivity several orders of magnitude smaller than standard soda-lime float glass. The glass, *model S8900*, is available from Schott Glass Technologies Inc. [6]. The gas gap of these chambers was also 1.15 mm and the area of the chambers measured $20 \times 20$ cm$^2$. With 1.4 mm thickness, the glass plates were somewhat thicker than for the other designs.

The main characteristics of the three types of chambers are summarized in Table I. For each type, two chambers were built, commissioned, and tested, for a total of six chambers exposed to the test beam.

**Table I.** Summary of the main features of the three different RPC designs.

| RPC design | Number of glass plates | Area $A$ [cm$^2$] | Bulk resistivity $\rho$ [$\Omega$cm] | Total thickness $t$ of the glass [cm] | Conductance per area of the glass $G = (\rho \cdot t)^{-1}$ [$\Omega^{-1}$cm$^{-2}$] | Rate at 50% efficiency [Hz/cm$^2$] |
|---|---|---|---|---|---|---|
| 1 | 2 | 400 | $4.7 \times 10^{12}$ | 0.22 | $1.0 \times 10^{-12}$ | 300 |
| 2 | 1 | 1536 | $3.7 \times 10^{12}$ | 0.11 | $2.4 \times 10^{-12}$ | 1500 |
| 3 | 2 | 400 | $6.3 \times 10^{10}$ | 0.28 | $5.6 \times 10^{-11}$ | 15,000 |

All chambers were flushed with the same mixture of three gases, R134a (94.5%), Isobutane (5.0%) and SF$_6$ (0.5%). The high voltage was chosen such that the chambers operated at the lower edge of the efficiency plateau of the avalanche mode with default high voltages in the range of 6.7 to 7.1 kV. One chamber only required 6.1 kV. The readout of the chambers utilized the electronic readout system [7] of the Digital Hadron Calorimeter, the DHCAL [8]. The system records hits above a single threshold, corresponding to a signal charge of approximately 110 fC. The pad board, located on the anode side of the chambers contained 1536 $1 \times 1$ cm$^2$ pads and thus was oversized for the smaller chambers. The data acquisition recorded the time stamps of each hit with a resolution of 100 ns. The data was acquired in either triggered mode (subsequent to an external trigger) or in triggerless mode, where every hit was recorded. The latter was utilized to monitor the performance of the system and to estimate the noise rate in the chambers.

## 3. Estimates of the noise rate

The accidental noise rate in the chambers was estimated using two different methods: a) for each trigger a sequence of hits corresponding to seven 100 ns time bins was recorded. Of these, two occurred before the arrival of the particles in the chambers and therefore can be used to estimate the accidental noise rate; b) during beam off conditions data were acquired in 60 second lasting triggerless runs. The two methods provided consistent results, as summarized in Table II, and show a slight increase in noise rate with decreasing overall resistance of the glass plates. The statistical errors of these measurements are small compared to the changes in rate due to changes in environmental conditions [9].

**Table II.** Summary of the accidental noise hit rate.

| RPC design | Number of glass plates | Conductance per area of the glass $G = (\rho \cdot t)^{-1}$ [$\Omega^{-1}$ cm$^{-2}$] | Noise rate from triggered events [Hz/cm$^2$] | Noise rate from triggerless mode [Hz/cm$^2$] |
|---|---|---|---|---|
| 1 | 2 | $1.0 \times 10^{-12}$ | 3.3 | 0.9 |
| 2 | 1 | $2.4 \times 10^{-12}$ | 1.6 | 1.6 |
| 3 | 2 | $5.6 \times 10^{-11}$ | 5.5 | 3.2 |

## 4. Set-up in the test beam and measurement strategy

The rate capability of the six RPCs was measured using the 120 GeV primary proton beam of the Fermilab Test Beam Facility FTBF [10]. As part of the standard equipment of the beam line three scintillator counters are placed upstream of the test area and provide a reliable measurement of the particle rate. In addition, three wire chambers can be used to determine both the horizontal and vertical sizes of the beam.

The six RPCs were placed downstream of the wire chambers and were spaced about 5 cm apart from each other in beam direction. The data acquisition was triggered with a set of two finger counters with a common cross section of $1 \times 1$ cm$^2$ and placed directly in front of the first RPC.

Due to a built-in dead time of the data acquisition of 0.3 ms following each trigger [7], at high beam intensities the data acquisition rate was significantly lower than the trigger rate. This, however, did not impact the determination of the chambers performance parameters. To first order, the efficiency of a given chamber was derived from the ratio of the number of events containing at least one hit to the total number of triggers accepted by the data acquisition. The average pad multiplicity was calculated using only events which contained at least one hit and is thus defined as an independent measure from the chambers efficiency.

## 5. Measurement of the beam intensity

The beam intensity was measured simultaneously by three scintillation counters placed in the beam. The particles arrived in spills spaced one minute apart and lasting approximately four seconds. An effort was made to extract the beam uniformly during the spill. As measured by the wire chambers, Fig. 1 shows the number of particles versus time within the spills. The left plot is an example of a low intensity run (30 Hz/cm$^2$), whereas the right plot is taken from a high intensity run (30 kHz/cm$^2$). In low intensity runs, the beam flux increased as the spill progressed, causing some uncertainty in the definition of the beam intensity for a given run. For high intensity runs the ambiguity is significantly smaller. In order to quantify the uncertainty in beam intensity due to the variation over the duration of a spill, the root mean square of the projection of the number of events onto the y-axis was assigned as systematic error.

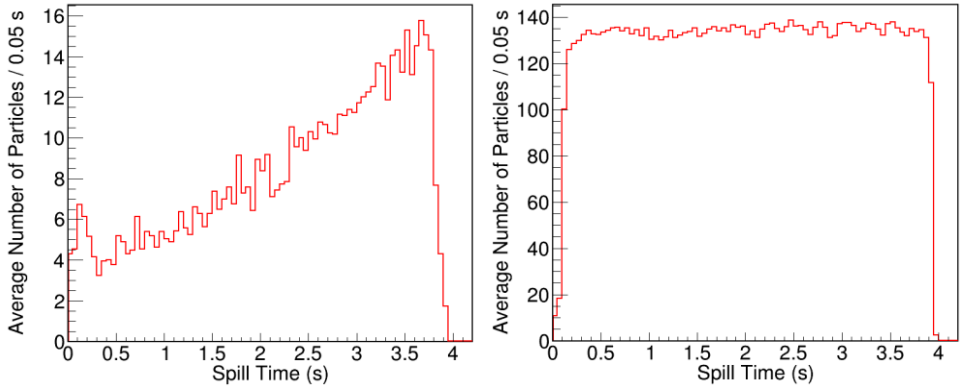

**Figure 1.** Number of events as function of time during the spill for a low intensity run (left) and a high intensity run (right).

The measurement of the rate capability in units of [$Hz/cm^2$] requires the knowledge of the size of the beam spot. In order to achieve a more accurate measurement of the beam spot, the usually pencil-like 120 GeV primary proton beam was defocused upstream of the experimental hall. The resulting beam profile was Gaussian in both horizontal and vertical direction with a width σ, as measured by the wire chambers, of approximately 1.0 (0.8) cm in the horizontal (vertical) direction. In the calculation of the beam intensity, in units of [$Hz/cm^2$], the size of the beam spot was taken to be $2\sigma_x \times 2\sigma_y$, with an error derived from the measurement error of the widths of the Gaussians.

In the subsequent analysis, the systematic errors due to the variation in beam intensity during the spill (dominant) and the uncertainty in the size of the beam spot were added in quadrature.

## 6. Measurement of the rate capability

In order to establish the rate capability of the chambers, the hits in each chamber and in each event were clustered using a nearest-neighbor clustering algorithm, requiring a common side for two hits to belong to the same cluster. The efficiency of the three sets of chambers was measured as the ratio of the number of events with a cluster within 3 cm of the average beam spot to the total number of triggered events. The error on the efficiency was calculated as a binomial error. The average pad multiplicity was defined as the average number of pads hit for events where that chamber recorded at least one hit.

In the present measurements, the initial exponential decrease in efficiency at the beginning of the spill, as reported in [2], was folded into the calculation of the efficiency. Since the spills are relatively long, the measurements of the efficiency are dominated by the later part of the spill where the efficiency is seen to be constant.

For all six chambers tested, Fig. 2 shows the efficiency (left) and the average pad multiplicity (right) as function of beam rate. As the rate increases the efficiency is observed to drop in all three sets of chambers. However, the loss of efficiency is shifted to higher rates for chambers with lower overall resistance of the glass. The average pad multiplicity is seen to be below two for the 2-glass chambers and constant and close to unity for the 1-glass chambers, as expected [5]. For the 2-glass chambers, there is an indication of a small rise in average pad multiplicities with increasing rates, which might be related to the higher probability for detecting protons which have interacted upstream of the chambers.

## 7. Conclusions

The rate capability of three sets of RPCs with different conductance per area of their glass plates has been measured. The results show that raising the overall conductance per area of the glass plates will enhance the rate capability and increase the range of particle rates for which the chambers retain their full particle detection efficiency. Figure 3 shows the rate $I_{50\%}$ at which the chamber perform with a 50% efficiency versus the conductance per area of the glass plates. The data was fit empirically to the following functional form

$$I_{50\%} = a + bH + cH^3 \qquad (1)$$

where $H = 1/\log_{10}(G)$, where $G$ is the conductance per area of the glass plates (as listed in Table I) and $a$, $b$, and $c$ are free parameters. The following values were obtained for the three free parameters of the function: $a = 1.7 \times 10^5$, $b = 3.2 \times 10^6$ and $c = -1.7 \times 10^8$.

Long term tests to establish possible aging effects of the 1-glass design and of the semi-conductive glass plates have begun and will be reported on in a follow-up paper.

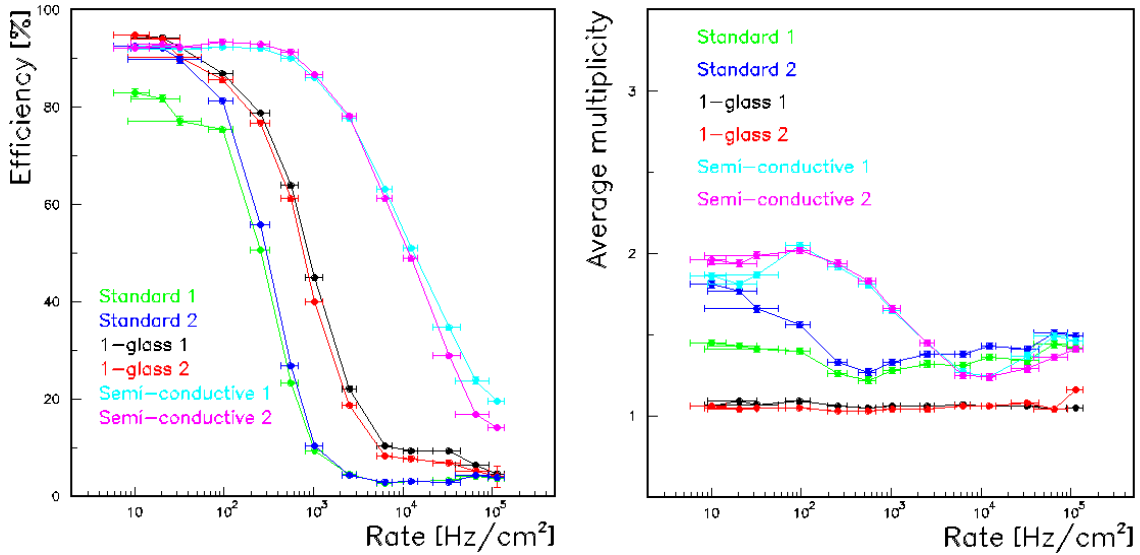

**Figure 2.** Efficiency (left) and average pad multiplicity (right) as function of beam rate for six different RPCs. For better visibility lines are drawn connecting the data points.

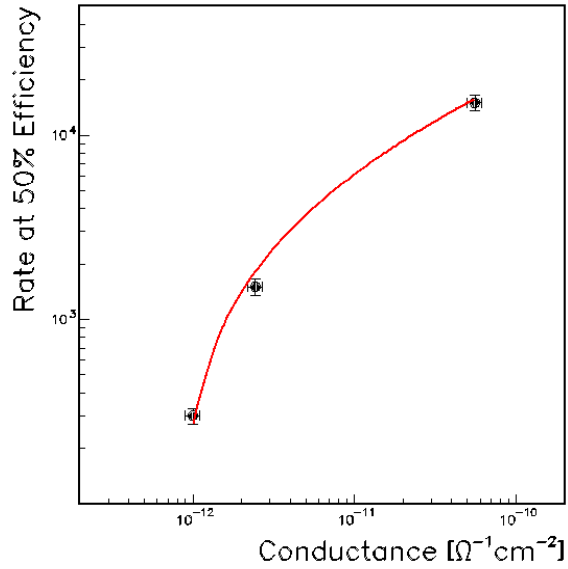

**Figure 3**. Rate at 50% efficiency versus conductance per area of the glass plates. The points were fitted to the function in Eq. (1). The result of the fit is shown as red line.